\documentclass[final,1p,times]{elsarticle} 

\usepackage{amsmath,amssymb}
\usepackage{color}

\journal{Nuclear Physics A} 

\begin{document}
\begin{frontmatter}

\title{Extracting the shear viscosity of the quark-gluon plasma from flow in ultra-central heavy-ion collisions}

\author[CNRS,mcgill,lbnl]{Matthew Luzum}
\author[CNRS]{Jean-Yves Ollitrault}


\address[CNRS]{
Institut de physique th\'eorique de Saclay (CNRS URA2306), F-91191
Gif-sur-Yvette, France
}

\address[mcgill]{
McGill University,  3600 University Street, Montreal QC H3A 2TS, Canada
}

\address[lbnl]{
Lawrence Berkeley National Laboratory, Berkeley, CA 94720, USA
}

\begin{abstract}
We propose a method for extracting the shear viscosity over entropy density
ratio ($\eta/s$) of the quark-gluon plasma from experimental data. 
We argue that uncertainty due to poor knowledge of the
earliest stages of a heavy-ion collision is smallest for ultra-central
events.  The most precise value of $\eta/s$ can thus be obtained from a global fit to $p_T$-integrated Fourier
harmonics of azimuthal correlations. 
We further outline a method for quantifying the overall uncertainty in
the extracted value.
 Only after a comprehensive and systematic accounting of 
all sources of 
uncertainty can a reliable measurement be claimed.
In these proceedings we report preliminary results; full
and final results will be presented in a separate publication. 
\end{abstract}

\end{frontmatter}

\section{Introduction}
Transport coefficients such as shear viscosity represent fundamental properties of a system and contain direct information about the underlying physics.  For QCD matter in the temperature range probed by an ultra-relativistic heavy-ion collision, such quantities are exceptionally difficult to calculate from first principles.
On the other hand, observables in these experiments can be quite sensitive to transport coefficients, 
potentially allowing for a measurement to be extracted from data~\cite{Alver:2010dn}.  
Thus, such a precision extraction is a topic of significant interest.

In the past, a precision extraction has been hampered by significant theoretical uncertainty, especially that arising from poor knowledge of particle production and thermalization in the earliest stages.  A number of viscous hydrodynamic calculations have been performed, which have been quite successful at reproducing flow measurements~\cite{Luzum:2008cw,Schenke:2011bn}.  Typically, a small value of viscosity is required in order to fit data---not too much more than a few times the conjectured lower bound
$\eta/s \geq 1/4\pi$.  However, the precise best-fit value 
depends on various aspects of the calculations that are not fully constrained, and this introduces uncertainty in any extraction 
from data.   Although these uncertainties have been the target of significant investigation, it has so far been done in a piecemeal fashion~\cite{Song:2008hj}.  No one has yet attempted to compile this information in a systematic study of all sources of uncertainty in order to place a robust and reliable upper bound on 
$\eta/s$.  In addition, the conjectured bound is now known not to be universal, so it is of great interest to determine if one can place a lower bound on the quantity.  Currently, there is no convincing evidence that measured data can not be reproduced with ideal 
hydrodynamics, and in fact most flow data can be reproduced quite well in such calculations~\cite{Gardim:2012yp}.  In these proceedings we present a proposal for how to extract $\eta/s$ with the smallest uncertainty currently possible, as well as preliminary results on such an extraction including a quantitative error bar obtained by systematic study of all known sources of uncertainty.
\section{Ultra-central collisions}
\begin{figure}
\includegraphics[width=0.99\linewidth]{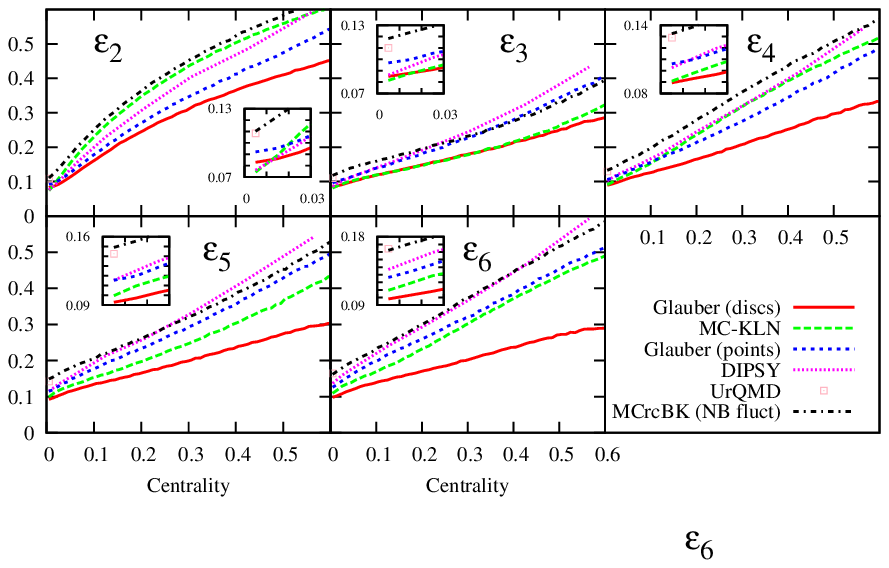}
\caption{(Color online) RMS eccentricities 
versus centrality for several models of initial conditions in 1\% centrality bins. 
Centrality was selected according to multiplicity or entropy, as appropriate for each model, and the transverse density of the same quantity was used to calculate eccentricities.  
Inset plots magnify the most central 3\% of collisions.  
The models shown are: PHOBOS MC-Glauber~\cite{Alver:2008aq}
[Glauber (points)], MC-KLN~\cite{Drescher:2007ax} v3.43 and its Glauber implementation [Glauber (discs)], mckt-v1.25 with negative binomial fluctuations~\cite{Dumitru:2012yr} (MCrcBK (NB)), DIPSY~\cite{Flensburg:2011wx}, and UrQMD~\cite{Petersen:2010cw}.}
\label{ecc}
\end{figure}
In a standard picture of a heavy-ion collision, the asymmetric region of overlap in a collision between two nuclei results in a  anisotropy in the transverse density of the system at early times.  Over time, this spatial anisotropy is transformed into a momentum anisotropy by a collective response 
that is well described by (viscous) hydrodynamics.  In fact, the elliptic flow coefficient $v_2$ is (to a good approximation) directly proportional to this initial eccentricity in a given collision event: 
$v_2 \propto \varepsilon_2 \equiv |\langle r^2 e^{2i\phi}
\rangle|/\langle r^2 \rangle$.  The proportionality constant depends
sensitively on the viscosity of the medium, 
However, poor knowledge of the earliest stages of the collision result
in significant uncertainty in $\varepsilon_2$
See, for example, the top left panel of Fig.~\ref{ecc}, which shows root-mean-square values of $\varepsilon_2$ from a number of commonly-used models for initial conditions\footnote{Most $v_n$ measurements  are sensitive to the rms value of the event-by-event distribution of $v_n$}.
The resulting 
uncertainty in $v_2$ has been the largest source of uncertainty when determining $\eta/s$ 
from data~\cite{Luzum:2008cw}.  Though there has been much recent study concerning the physics of the early stages of a collision~\cite{Dumitru:2012yr,Flensburg:2011wx}, 
it is not yet possible to make a reliable statement about what range
of values of initial eccentricity are within the realm of possibility.  

It is now known that quantum fluctuations are also a significant source of anisotropy in the initial state, and therefore anisotropic flow.  This is especially apparent from the presence of large odd harmonics, such as triangular flow $v_3$, which is generated from a spatial ``triangularity'', defined similarly to the eccentricity
$v_3 \propto \varepsilon_3$
This would be zero at midrapidity in the absence of event-by-event fluctuations.  
The 
triangularity that one sees in a calculation is largely determined by the strength of these fluctuations.  In contrast, the differences in particle production mechanisms that lead to very different $\varepsilon_2$ in non-central collisions has very little effect on $\varepsilon_3$.  Compare, for example, the triangularities of the MC-KLN model versus Glauber (discs) in Fig.~\ref{ecc}, which contain only fluctuations from random position of nucleons in the colliding nuclei, implemented in exactly the same way in the MC-KLN code.  The triangularity $\varepsilon_3$ is very similar, despite the fact that the two pictures of particle production result in a very different eccentricity $\varepsilon_2$.  

By comparing the results from other models, it is clear that there still remains some uncertainty due to different way fluctuations are implemented.  Despite the seemingly modest reduction in uncertainty (as compared to the uncertainty in $\varepsilon_2$), one can reasonably argue that the results in the second panel of Fig.~\ref{ecc} represent the full range of results that can be expected, in marked contrast to the results in the first panel.  The models with the lowest triangularity are known to be missing important sources of fluctuations, while the large fluctuations in the models with the largest $\varepsilon_3$ cannot be reconciled with constraints from $v_1$ measurements~\cite{Retinskaya:2012ky}.

One might thus propose to extract $\eta/s$ from measurements of $v_3$ to minimize uncertainty from the initial conditions.  However, we can make use of more experimental constraints by noting that in the most central collisions, all harmonics are generated entirely from fluctuations, and the above discussion still applies.  One further convenient factor is that, for ultra-central collisions, the approximate proportionality $v_n\propto\varepsilon_n$ holds for all $n$~\cite{Gardim:2011xv}.  This allows for a comprehensive parameter study with reasonable computing resources.

Thus, our proposal is to extract $\eta/s$ from a simultaneous fit to $p_T$-integrated $v_n$ measurements from ultra-central collisions at the highest-energy heavy-ion collisions.  These data are least sensitive to uncertain aspects of theory, and therefore allow for the the most precise extraction of $\eta/s$ with the smallest possible uncertainty.  Such data are available from the ATLAS collaboration for $v_2$--$v_6$ in the 1\% of most central collisions~\cite{ATLAS:2012at}.  

\begin{figure}
\includegraphics[width=0.49\linewidth]{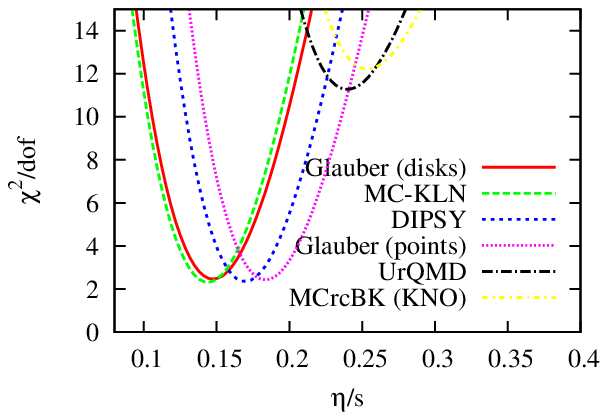}
\includegraphics[width=0.49\linewidth]{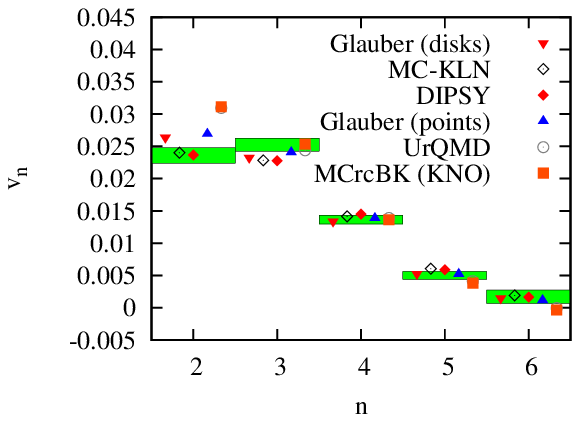}
\caption{(Color online) 
Left:  Example plot of chi-squared per degree of freedom for a one-parameter simultaneous fit of integrated $v_n$ for n = 2--6 at 0--1\% centrality, as a function of the ratio of shear viscosity to entropy density $\eta/s$.  Right:  Fit values corresponding to the minimum chi-squared curves in the left plot.  The green bars represent measured values from the ATLAS collaboration~\cite{ATLAS:2012at}; the vertical thickness represents the combined statistical and systematic error.  Models with large fluctuations and a correspondingly large best-fit $\eta/s$ tend to have a steeper slope than models with smaller eccentricities.  The range of best-fit viscosity values for the curves in the left plot gives an indication of the uncertainty in an extraction of $\eta/s$ due to the uncertain initial anisotropy in the initial state ($\sim \pm$ 0.05).
}
\label{chi}
\end{figure}
\section{Extracting $\eta/s$}
In a theoretical calculation, the viscosity of the medium can be taken to be a free parameter.  With all other parameters fixed, one can calculate $v_n$ as a function of $\eta/s$, and perform a least-squares fit to find the preferred value.  If a particular parameter is uncertain, one can vary that parameter within the allowed range and redo the fit.  The shift of best-fit $\eta/s$ indicates the size of the resulting error bar in $\eta/s$ due to this particular source of uncertainty.  If one then varies all possible aspects of the theory in all possible combinations, the smallest and the largest extracted values thus represent a robust, if conservative, upper and lower error bar for the quantity.

This is the idea behind the preliminary results shown at this conference.  Details will be given in a forthcoming publication, but an example fit for a calculation using a single set of hydrodynamic parameters but different initial conditions is shown in Fig.~\ref{chi}.  It is notable that hydrodynamics can, in fact, simultaneously give a good description of all $v_n$.  However, it is also interesting to note the tendency for tension in the relative values of $v_2$ and $v_3$ for all initial conditions tested.

We reported a preliminary result of $0.07\leq\eta/s \leq 0.43$, but a few comments are in order.  First, the calculations used a constant value of $\eta/s$ during the hydrodynamic evolution.  Therefore, the result is an average value that corresponds to the viscosity at the typical time when flow is built up.  The lower bound indicates that purely ideal hydrodynamic calculations can not describe these data.  This is a significant result.   However, it is technically possible that the viscosity could dip below our lower bound at some point during the evolution, so this result does \textit{not} represent a lower bound for the viscosity of QCD over its entire temperature range.  Second, since it has never before been done, we want to be confident at every stage that we have a robust and reliable error bar.  The preliminary result, therefore, is fairly conservative, and the error bar will likely be smaller in the final result after further study.
\section{Conclusions}
We argue that at the current time, the most precise value for the shear viscosity to entropy density ratio $\eta/s$ of the quark-gluon plasma can be obtained by a fit to $p_T$-integrated $v_n$ measurements in ultra-central collisions.   We further report preliminary results such an extraction, including an error bar from a systematic study of all known sources of uncertainty.

We thank the PHOBOS Collaboration, Adrian Dumitru, Yasushi Nara, and Christoffer Flensburg for providing their code to the public, which allowed us to calculate the eccentricities in various Montre-Carlo models, and Hannah Petersen for calculating them for us using UrQMD.


\end{document}